\documentclass[twocolumn,preprintnumbers,elsart]{revtex4-1}
\usepackage{color}
\usepackage{eurosym}
\usepackage{makeidx}
\usepackage{amssymb}
\usepackage{amsmath}
\usepackage{mathrsfs}
\usepackage{graphicx}
\usepackage{dcolumn}
\usepackage{bm}
\usepackage{makeidx}

\begin{document}

\title{Singular solitons}
\author{Hidetsugu Sakaguchi$^{1}$ and Boris A. Malomed$^{2}$}
\affiliation{$^{1}$Department of Applied Science for Electronics and Materials,
Interdisciplinary Graduate School of Engineering Sciences, Kyushu
University, Kasuga, Fukuoka 816-8580, Japan\\
$^{2}$Department of Physical Electronics, School of Electrical Engineering,
Faculty of Engineering, and Center for Light-Matter Interaction, Tel Aviv
University, Tel Aviv 69978, Israel}

\begin{abstract}
We demonstrate that the commonly known concept, which treats solitons as
nonsingular solutions produced by the interplay of nonlinear self-attraction
and linear dispersion, may be extended to include modes with a relatively
weak singularity at the central point, which keeps their integral norm
convergent. Such states are generated by self-repulsion, which should be
strong enough, namely, represented by septimal, quintic, and usual cubic
terms in the framework of the one-, two-, and three-dimensional (1D, 2D, and
3D) nonlinear Schr\"{o}dinger equations (NLSEs), respectively. Although such
solutions seem counterintuitive, we demonstrate that they admit a
straightforward interpretation as a result of screening of an additionally
introduced attractive delta-functional potential by the defocusing
nonlinearity. The strength (\textquotedblleft bare charge") of the
attractive potential is infinite in 1D, finite in 2D, and vanishingly small
in 3D. Analytical asymptotics of the singular solitons at small and large
distances are found, entire shapes of the solitons being produced in a
numerical form. Complete stability of the singular modes is accurately
predicted by the \textit{anti-Vakhitov-Kolokolov} criterion (under the
assumption that it applies to the model), as verified by means of numerical
methods. In 2D, the NLSE with a quintic self-focusing term admits
singular-soliton solutions with intrinsic vorticity too, but they are fully
unstable. We also mention that dissipative singular solitons can be produced
by the model with a complex coefficient in front of the nonlinear term.
\end{abstract}

\maketitle

\section{Introduction}

Many physical media naturally feature competing nonlinearities, which, in
turn, help to maintain specific soliton states in one-, two-, and
three-dimensional (1D, 2D, 3D) settings. In particular, a known result is
that, in the framework of the nonlinear Schr\"{o}dinger equations (NLSEs),
the combination of self-focusing cubic and defocusing quintic nonlinear
terms supports, in addition to fundamental 2D and 3D solitons, self-trapped
modes with embedded vorticity, which are stable against splitting. In the 2D
cubic-quintic (CQ) medium, vortex solitons have their stability regions for
all values of the winding number, $S$, which shrink with the increase of $S$
\cite{Manolo,Bob}. Similar predictions were made for two-component CQ
systems \cite{two-comp}. In 3D, a stability region was identified for
donut-shaped solitons with embedded vorticity $S=1$, using both CQ \cite{3D}
and quadratic-cubic \cite{quadr-cubic} combinations of competing nonlinear
terms (see also a review in Ref. \cite{PhysD}). Stable 3D soliton clusters
in the CQ medium were predicted too \cite{3D-cluster}. A recent addition to
the topic is the derivation of the 3D Gross-Pitaevskii equation (GPE)\ for a
binary Bose-Einstein condensate (BEC), with the cubic self-attraction and
quartic repulsion, which represents effects of quantum fluctuations around
the mean-field states \cite{Petrov}. The reduction of the latter setting to
2D leads to the single nonlinear term, in the form of the cubic term
multiplied by a logarithmic factor \cite{Astra}. This prediction was
followed by experimental creation of stable soliton-like states without
vorticity, in the form of ``quantum droplets" \cite%
{Leticia1,Leticia2,hetero,Inguscio1}. The condensate filling such states is
considered as an ultradilute quantum fluid, as the interplay of the
self-focusing and defocusing terms makes it essentially incompressible, with
the density which cannot exceed a particular maximum value. Accordingly, the
increase of the condensate's norm lends the droplets a \textit{flat-top}
shape (called, for this reason, ``quantum puddles" in Ref. \cite{AM}).
Further theoretical work has predicted stable 3D droplets with embedded
vorticities $S=1$ and $2$ \cite{Yaro}, and 2D ring-shaped droplet clusters
\cite{Yaro2}, as well as stable 2D vortex droplets with $1\leq S\leq 5$, as
well as ``hidden-vorticity" states, in which two components of the binary
condensate have equal norms and opposite values of the angular momentum \cite%
{2D-vort-drop}.

CQ nonlinearity is known in various optical materials, which has made it
possible to create stable fundamental ($S=0$)\ solitons in an effectively 2D
form \cite{Brazil}. In the same setting, 2D solitons with embedded vorticity
$S=1$ were created as long-lived transient modes, stabilized with the help
of cubic loss \cite{transient}. Optical media with controlled strengths of
cubic, quintic, and even septimal terms can be realized in suspensions of
metallic nanoparticles, using the density and size of the particles as
control parameters \cite{Cid,Cid2}. The creation of stable 2D spatial
solitons supported by the quintic-septimal nonlinearity (in the absence of
the cubic term) was reported too \cite{Cid3}.

In the normalized 1D form, the cubic-quintic-septimal NLSE for the local
amplitude of the electromagnetic wave, $u$, propagating in the spatial
domain (i.e., in a planar waveguide with longitudinal and transverse
coordinates, $z$ and $x$), is
\begin{equation}
iu_{z}=-\frac{1}{2}u_{xx}+g_{3}|u|^{2}u+g_{5}|u|^{4}u+|u|^{6}u,  \label{NLSE}
\end{equation}%
where the coefficient in front of the self-defocusing septimal term is
scaled to be $1$, while $g_{3}$ and $g_{5}$ account for the CQ nonlinearity
\cite{Reyna,Cid}. A commonly known principle is that the presence of
self-focusing is necessary for the existence of bright solitons \cite%
{KA,Peyrard,Yang}. In terms of Eq. (\ref{NLSE}), this necessary condition
amount to $g_{3}<0$, or $g_{3}>0$ and $g_{5}<-2\sqrt{g_{3}}$, the largest
possible amplitude, corresponding to the above-mentioned flat-top solitons,
being
\begin{equation}
u_{\max }=\sqrt{\frac{1}{3}\left( \sqrt{g_{5}^{2}-3g_{3}}-g_{5}\right) }.
\end{equation}

The objective of the present work is to demonstrate that the self-defocusing
septimal term in 1D equation (\ref{NLSE}) supports a family of bright
solitons with an \emph{integrable singularity} $\sim |x|^{-1/3}$ at $%
x\rightarrow 0$, i.e., the norm of the singular soliton converges (in terms
of the above-mentioned realization in optics, the norm is tantamount to the
total power of the light beam). On the contrary, if the leading repulsive
nonlinear term is quintic or cubic, it gives rise to formal 1D soliton
solutions with singularities $\sim |x|^{-1/2}$ and $|x|^{-1}$, respectively,
which are unphysical states with divergent norms. We elaborate a physical
interpretation of this counter-intuitive finding, demonstrating that it may
be construed as a result of screening by the septimal repulsive nonlinearity
of an attractive delta-functional potential with an infinite strength. The
entire family of the singular 1D solitons satisfies the necessary stability
condition in the form of the \textit{\ anti-Vakhitov-Kolokolov} (aVK)
criterion \cite{we} (although the applicability of the aVK criterion to the
present model is only a conjecture), and is indeed stable, as demonstrated
by numerical results.

Further, the 2D version of NLSE with the leading quintic repulsive term [see
Eq. (\ref{u2D}) below, with $\sigma =1$] supports bright solitons with
singularity $\sim r^{-1/2}$ ($r$ is the radial coordinate) and convergent 2D
norm, which are stable too, according to the aVK criterion and numerical
results alike. Its physical interpretation is also available in the form of
screening of a singular attractive potential by the quintic self-repulsion,
but the difference from the 1D case is that the strength of the 2D singular
potential is finite.

It is relevant to mention that the existence of solutions of the stationary
nonlinear 1D and 2D equations with the above-mentioned singularities was
known before \cite{Veron1}, see also Ref. \cite{Veron2}. However, the global
shape of the solutions, their stability, in the framework of the NLSE, and
the physical interpretation, in terms of the screened singular potentials,
were not considered previously.

In addition, we demonstrate that the 2D model with the \emph{attractive}
quintic term [$\sigma =-1$ in Eq. (\ref{u2D})] gives rise to singular
solitons with embedded vorticity, but they are completely unstable. The
attractive nonlinearity was not considered in Ref. \cite{Veron1}.

In the 3D geometry, stable isotropic bright solitons with singularity $\sim
r^{-1}\left( \ln \left( r_{0}/r\right) \right) ^{-1/2}$ and convergent norm
are created by the usual repulsive cubic term, the respective equation (\ref%
{u3D}) (see below) being the usual GPE for the BEC in the mean-field
approximation (this case was not considered in Ref. \cite{Veron1} either).
The interpretation of the singular 3D solitons in terms of the screening of
a singular attractive potential is possible too, with a conclusion that the
strength of the potential is vanishingly small.

Lastly, it is relevant to mention that the cubic repulsive nonlinearity may
support stable non-singular 1D, 2D, and 3D solitons, self-trapped vortices
\cite{Barcelona1}, and even stable 3D \textit{hopfions} \cite{Barcelona2},
by means of a completely different mechanism, if the local strength of the
self-repulsion grows fast enough from the center to periphery.

The rest of the paper is organized as follows. Analytical and numerical
results are reported, severally, in Sections 2 and 3, each one split into
subsections according to the spatial dimension. In particular, a possibility
of having dissipative singular solitons in the model with a complex
nonlinearity coefficient is briefly discussed in Section 2. The paper is
concluded by Section 4, which also briefly discusses a possibility of the
experimental implementation of the results in optics and BEC.

\section{Analytical considerations}

\subsection{The 1D model}

\subsubsection{The singular soliton created by the septimal nonlinearity}

The most essential results for the 1D case can be produced by the
septimal-only equation (\ref{NLSE}), with $g_{3}=g_{5}=0$. This model may be
realized in the planar optical waveguide based on the colloidal medium, by
adjusting its parameters \cite{Cid2}. As shown at the end of the present
subsection, the inclusion of the cubic and quintic terms produces an
insignificant effect on results presented below.

Stationary solutions to Eq. (\ref{NLSE}) with propagation constant $k>0$ are
looked for as
\begin{equation}
u\left( x,z\right) =\exp \left( ikz\right) U(x),  \label{uU}
\end{equation}%
where real function $U(x)$ satisfies equation%
\begin{equation}
\frac{1}{2}\frac{d^{2}U}{dx^{2}}=kU+U^{7},  \label{U}
\end{equation}%
with the respective (formal) Hamiltonian,%
\begin{equation}
h=\left( \frac{dU}{dx}\right) ^{2}-2kU^{2}-\frac{1}{2}U^{8}.  \label{h}
\end{equation}

Using Eq. (\ref{h}) with $h=0$, that corresponds to solitons, the solution
can be found in an implicit form, which treats the coordinate as a function
of $U$:
\begin{equation}
x(U)=\frac{1}{2\sqrt{2k}}\int_{U^{2}/\left( 4k\right) ^{1/3}}^{\infty }\frac{%
dW}{W\sqrt{1+W^{3}}}.  \label{x}
\end{equation}%
At $x\rightarrow 0$, the universal ($k$-independent) asymptotic form of the
singular soliton is%
\begin{equation}
U(x)\approx \left( 2/9\right) ^{1/6}|x|^{-1/3}  \label{-1/3}
\end{equation}%
(as mentioned above, this asymptotic form was first found in Ref. \cite%
{Veron1}), and in the limit of $|x|\rightarrow \infty $ its exponentially
decaying tail is%
\begin{equation}
U(x)\approx U_{0}\exp \left( -\sqrt{2k}|x|\right) ,  \label{tail}
\end{equation}%
with constant $U_{0}$. An approximate global form of the soliton can be
produced by matching the asymptotic expressions (\ref{-1/3}) and (\ref{tail}%
):%
\begin{widetext}
\begin{equation}
U(x)\approx \left\{
\begin{array}{c}
\left( \frac{2}{9}\right) ^{1/6}|x|^{-1/3}\,,~\mathrm{at}~~|x|\leq \frac{1}{3%
\sqrt{2k}}, \\
\left( 2\sqrt{k}\right) ^{1/3}\exp \left( -\sqrt{2k}\left( |x|-\frac{1}{3%
\sqrt{2k}}\right) \right) ,~\mathrm{at}~~|x|\geq \frac{1}{3\sqrt{2k}},%
\end{array}%
~\right.   \label{match}
\end{equation}
\end{widetext}where the matching point, $|x|=1/\left( 3\sqrt{2k}\right) $,
is selected as one at which both $U(x)$ and its first derivative are
continuous, as given by the two expressions in Eq. (\ref{match}). As shown
below in Fig. \ref{f1}(b), Eq. (\ref{match}) provides a virtually exact form
of the 1D singular solitons.

The use of the septimal nonlinearity is essential here, because a general
leading self-repulsive nonlinear term, $\sim |u|^{2n}u$, will generate a
solution with $U(x)\sim |x|^{-1/n}$ at $x\rightarrow 0$, hence for $n=1$
(cubic) and $n=2$ (quintic) terms, the singular density is, respectively, $%
U^{2}\sim |x|^{-2}$ and $U^{2}\sim |x|^{-1}$, making the total norm
divergent. On the other hand, the norm of the soliton, given in the implicit
form by Eq. (\ref{x}), converges because the singularity of $U^{2}(x)$ at $%
x\rightarrow 0$ is integrable, according to Eq. (\ref{-1/3}):%
\begin{equation}
N=\int_{-\infty }^{+\infty }U^{2}(x)dx=3.1478\cdot k^{-1/6},  \label{N}
\end{equation}%
where the numerical coefficient is
\begin{equation}
2^{1/6}\int_{0}^{\infty }\frac{dx}{\sqrt{1+x^{3}}}\approx \allowbreak
3.\allowbreak 1478.
\end{equation}%
The approximate form (\ref{match}) of the solution yields a result similar
to Eq. (\ref{N}), with coefficient $2^{7/6}+2^{1/6}=\allowbreak
3.\,\allowbreak 367\,4$, instead of $3.1478$, the relative error being $%
\simeq 0.07$.

In the case of $k=0$, solution (\ref{-1/3}) is an \emph{exact} one, but its
total norm diverges at large values of $|x|$, as is also seen from Eq. (\ref%
{N}), setting $k=0$ in it. Note that the $N(k)$ dependence, given by Eq. (%
\ref{N}), satisfies the above-mentioned aVK criterion, $dN/dk<0$, which is
conjectured to be a necessary stability condition for solitons supported by
self-repulsive nonlinearities
\index{we}.

Lastly, if the quintic and cubic terms are present in the underlying 1D
equation (\ref{NLSE}), it is easy to see that they produce negligible
corrections to the asymptotic singular form (\ref{-1/3}) at $x\rightarrow 0$:%
\begin{equation}
U_{1}(x)\approx -\left(
\frac{32}{81}\right) ^{1/6}\frac{g_{5}}{5}|x|^{1/3}-\frac{3g_{3}}{7\sqrt{2}}%
|x|.  \label{delta1D}
\end{equation}

\subsubsection{Interpretation of the singular soliton: screening of an
attractive $\protect\delta $-functional potential}

The counter-intuitive result demonstrating the existence of the 1D singular
solitons under the action of the septimal self-repulsion can be understood
by comparing Eq. (\ref{U}) with the following stationary equation:%
\begin{equation}
\frac{1}{2}\frac{d^{2}U}{dx^{2}}=kU+U^{7}-\varepsilon \delta (x)U,
\label{eps}
\end{equation}%
which includes the attractive delta-functional potential with large strength
$\varepsilon >0$. Indeed, at $x^{2}\ll 1/k$ (in the region where term $kU$
in Eq. (\ref{eps}) is negligible) the solution of Eq. (\ref{eps}) may be
approximated by a regularized version of expression (\ref{-1/3}):%
\begin{equation}
U_{\varepsilon }(x)\approx \left( \frac{2}{9}\right) ^{1/6}\left( |x|~+\xi
\right) ^{-1/3},  \label{approx}
\end{equation}%
with offset $\xi >0$ determined by the jump of the derivative of the
solution to Eq. (\ref{eps}) at $x=0$:%
\begin{equation}
\xi =1/\left( 3\varepsilon \right) .  \label{xi}
\end{equation}%
Note that the substitution of the approximation based on Eqs. (\ref{eps})
and (\ref{xi}) in the total Hamiltonian corresponding to Eq. (\ref{eps}),%
\begin{equation}
H=\int_{-\infty }^{+\infty }\left[ \frac{1}{2}\left( \frac{dU}{dx}\right)
^{2}+\frac{1}{4}|U|^{8}\right] dx-\varepsilon |U(x=0)|^{2},  \label{H}
\end{equation}%
yields
\begin{equation}
H_{\varepsilon }\approx -\left( 1/5\right) \left( 3\varepsilon \right)
^{5/3}.  \label{Heps}
\end{equation}%
The negative sign of expression (\ref{Heps}) suggests (in addition to the
above-mentioned aVK criterion) that the respective solution is stable, as
the ground state of the Hamiltonian.

Comparison of numerical solutions of Eq. (\ref{eps}) for $\varepsilon =40$
and $400$ with the singular soliton, produced by Eqs. (\ref{U}), (\ref{x}),
and (\ref{-1/3}), is displayed below in Fig. \ref{f-extra}. It confirms that
the numerical solutions indeed approach the singular soliton with the
increase of $\varepsilon $.

Thus, the fact that, as per Eq. (\ref{xi}), the strength (``charge") of the
delta-functional attractive potential $\varepsilon $ diverges in the limit
of $\xi \rightarrow 0$, which corresponds to the exact solution given by
Eqs. (\ref{x})-(\ref{tail}), suggests a physical interpretation of the 1D
model: an \emph{infinitely large} point-like ``charge" embedded in the
medium with the self-defocusing septimal nonlinearity is \emph{completely
screened} by the nonlinearity, which builds the singular soliton with the
convergent norm, for this purpose. This mechanism is somewhat similar to the
renormalization procedure in quantum electrodynamics, where an infinite
\textit{bare charge} of the electron may be cancelled by other diverging
factors, to produce finite physically relevant predictions, while the
electron remains a point-like singularity of the electric field \cite{QED}.

\subsection{The 2D model}

\subsubsection{Singular solitons generated by the quintic nonlinearity}

In the 2D model, the crucial role is played by the quintic term, the
respective equation being
\begin{equation}
iu_{z}=-\frac{1}{2}\nabla ^{2}u+\sigma |u|^{4}u,  \label{u2D}
\end{equation}%
where $\sigma =+1$ and $-1$ corresponds, respectively, to the self-repulsion
and self-attraction. As shown at the end of the present subsection, a cubic
term, if it is kept in Eq. (\ref{u2D}), produces an insignificant effect on
the solution. Similar to Eq. (\ref{NLSE}), this equation models the paraxial
propagation of the light beam in the bulk waveguide filled by the colloid
with parameters selected so as to strongly enhance the quintic term in the
dielectric response \cite{Cid,Cid2}.

In polar coordinates $\left( r,\theta \right) $, a solution to Eq. (\ref{u2D}%
) with embedded vorticity, $S=0,1,2,...$ is looked for as
\begin{equation}
u=\exp \left( ikz+iS\theta \right) U(r),  \label{M}
\end{equation}%
where real amplitude function $U(r)$ obeys the radial equation:%
\begin{equation}
kU-\frac{1}{2}\left( \frac{d^{2}U}{dr^{2}}+\frac{1}{r}\frac{dU}{dr}-\frac{%
S^{2}}{r^{2}}U\right) +\sigma U^{5}=0.  \label{U2D}
\end{equation}

For $S=0$ and $\sigma =+1$ (the repulsive nonlinearity), 2D fundamental
solitons with a singularity at $r\rightarrow 0$ have the form of
\begin{equation}
U_{\mathrm{2D}}(r)\approx 2^{-3/4}r^{-1/2}+2^{1/4}kr^{3/2},  \label{2Drep}
\end{equation}%
where the first correction $\sim k$ is kept too (as mentioned above, this
singularity was first found in Ref. \cite{Veron1}). The respective 2D norm, $%
N=2\pi \int_{0}^{\infty }U^{2}(r)rdr$, converges at $r\rightarrow 0$. For $%
k=0$,
\begin{equation}
U_{\mathrm{2D}}^{(k=0)}(r)=2^{-3/4}r^{-1/2}  \label{u0}
\end{equation}%
is an \emph{exact solution} of Eq.~(\ref{U2D}), but with the integral norm
diverging at $r\rightarrow \infty $.

For vortex states with $S\geq 1$, a similar solution with the integrable
singularity exists if the nonlinearity is \emph{self-attractive}, i.e., with
$\sigma =-1$ in Eq. (\ref{u2D}):
\begin{equation}
U_{\mathrm{2D}}^{(S)}(r)\approx \left[ \frac{1}{2}\left( S^{2}-\frac{1}{4}%
\right) \right] ^{1/4}r^{-1/2}  \label{2Dattr}
\end{equation}%
[in the case of $k=0$, Eq. (\ref{2Dattr}) yields an exact solution of Eq. (%
\ref{U2D}) with diverging norm]. On the contrary to usual vortex solutions
in which the amplitude vanishes at $r\rightarrow 0$, the present one
diverges in the same limit, which is an alternative form of the amplitude
profile compatible with the embedded vorticity (cf. the asymptotic forms of
the standard Bessel and Neumann functions at $r\rightarrow 0$).
Vortex-soliton solutions with a similar singularity were found in Ref. \cite%
{we2} in a 2D model combining the quintic self-repulsive term and an
attractive potential $\sim -r^{-2}$. Because the solution with $S\geq 1$
exists in the case of the self-attractive quintic nonlinearity, which drives
the \textit{supercritical collapse} in 2D \cite{Berge,Fibich}, it is
plausible that the vortex solutions are unstable. Indeed, it follows from
Eq. (\ref{U2D}) that the 2D norm of the solutions, with $S=0$ and $S\geq 1$
alike, obeys an \emph{exact} scaling relation,
\begin{equation}
N_{\mathrm{2D}}(k)=\mathrm{const}\cdot k^{-1/2},  \label{N2D}
\end{equation}%
where $\mathrm{const}$ depends on $S$. It follows from numerical results
reported below that $\mathrm{const}\approx 21.2$ for 2D solitons with $S=0$.
Dependence (\ref{N2D}) satisfies the aVK criterion, but \emph{not} the
Vakhitov-Kolokolov criterion per se, $dN/dk>0$ \cite{Vakh,Berge,Fibich},
which is a necessary stability condition in the case of an attractive
nonlinearity \cite{Vakh,Berge,Fibich}. This is an additional argument
predicting instability of the singular vortices, while the singular 2D
solitons with $S=0$ are expected to be stable as solutions of Eq. (\ref{u2D}%
) with self-repulsion ($\sigma =+1$). The expectations are corroborated
below by numerical results.

At $r\rightarrow \infty $, the asymptotic form of the solution is determined
by the linearization of Eq. (\ref{U2D}),%
\begin{equation}
U_{\mathrm{2D}}(r)\approx \frac{C}{\sqrt{r}}\left( 1-\frac{1}{8\sqrt{2k}r}%
\right) \exp \left( -\sqrt{2k}r\right)  \label{2Dexp}
\end{equation}%
[cf. Eq. (\ref{tail})] where $C$ is a constant, and the second term in the
parenthesis represents the first correction to the lowest-order
approximation. A global approximation, similar to its 1D counterpart (\ref%
{match}), which is produced by matching asymptotic forms (\ref{2Drep}) and (%
\ref{2Dexp}) at some intermediate point, demanding the continuity of $U_{%
\mathrm{2D}}(r)$ and its derivative, takes a cumbersome form in the 2D case,
therefore it is not written here.

Lastly, if the cubic term, the same as in Eq. (\ref{NLSE}), is added to Eq. (%
\ref{u2D}), it produces a negligible correction to the singular asymptotic
form given by Eq. (\ref{2Drep}) (for $\sigma =+1$): $U_{1}(r)\approx
-2^{-5/4}g_{3}\sqrt{r}$, cf. Eq. (\ref{delta1D}).

\subsubsection{Interpretation of the 2D singular soliton in terms of
screening of a singular attractive potential}

Similar to the interpretation of the 1D model, one can try to introduce, at
an intermediate stage, the 2D model with a delta-function, but this time
concentrated, instead of the single point, on a ring of a small radius, $%
\rho $. The respective stationary equation with $\sigma =+1$ (quintic
self-repulsion) and $S=0$ is [cf. Eqs. (\ref{U2D}) and (\ref{eps})]%
\begin{equation}
kU-\frac{1}{2}\left( \frac{d^{2}U}{dr^{2}}+\frac{1}{r}\frac{dU}{dr}\right)
+U^{5}-\varepsilon _{\mathrm{2D}}\delta (r-\rho )U=0.  \label{2Deps}
\end{equation}%
The solution is then looked for in the form given by Eq. (\ref{2Drep}) at $%
r>\rho $, and as%
\begin{equation}
U=\mathrm{const}\approx 2^{-3/4}\rho ^{-1/2}~~\mathrm{at}~~r<\rho .
\label{const}
\end{equation}%
Then, integration of Eq. (\ref{2Deps}) in an infinitely small vicinity of $%
r=\rho $ yields a jump condition for the radial derivative:%
\begin{equation}
\left( \frac{1}{U}\frac{dU}{dr}\right) |_{r=\rho +0}=-2\varepsilon _{\mathrm{%
2D}}.  \label{rho1}
\end{equation}%
On the other hand, the leading (first) term in solution (\ref{2Drep}) yields
its own value for the same quantity:
\begin{equation}
\left( \frac{1}{U}\frac{dU}{dr}\right) |_{r=\rho }=-\frac{1}{2}\rho ^{-1}.
\label{rho2}
\end{equation}%
Equating expressions (\ref{rho1}) and (\ref{rho2}) yields%
\begin{equation}
\varepsilon _{\mathrm{2D}}=1/\left( 4\rho \right) ,  \label{2D}
\end{equation}%
cf. Eq. (\ref{2D}). Thus, in the limit of $\rho \rightarrow 0$, an effective
total ``charge" $Q_{\mathrm{2D}}$ emulating the current solution is produced
by the integration of $\varepsilon _{\mathrm{2D}}$ over the ring of radius $%
\rho $:%
\begin{equation}
Q_{\mathrm{2D}}=2\pi \rho \cdot \varepsilon _{\mathrm{2D}}=\frac{\pi }{2}.
\label{Q2D}
\end{equation}%
This result suggests that the present 2D solution may be interpreted as a
result of screening of the ``charge" by the quintic nonlinearity, for the
finite value (\ref{Q2D}) of the charge.

\subsection{The 3D model}

\subsubsection{A singular soliton generated by the cubic self-repulsion}

In the 3D case, the crucial role is played by the usual cubic self-repulsive
term in the NLSE:
\begin{equation}
iu_{t}=-\frac{1}{2}\nabla ^{2}u+|u|^{2}u,  \label{u3D}
\end{equation}%
which has a well-known physical realizations in the 3D space \cite{review},
the most straightforward one being the GPE for the usual BEC with repulsive
interatomic interactions. Evolution variable $z$ in Eq. (\ref{u3D}) is
replaced by time $t$, which is relevant for the GPE, and, accordingly,
propagation constant $k$ is replaced by $-\omega $ below, cf. Eqs. (\ref{uU}%
) and (\ref{M})]. Thus, isotropic solutions are looked for as $u=\exp
(-i\omega t)U(r)$, where $r$ is the radial coordinate in the 3D space, and
real function $U$ satisfies the radial equation,
\begin{equation}
-\omega U-\frac{1}{2}\left( \frac{d^{2}U}{dr^{2}}+\frac{2}{r}\frac{dU}{dr}%
\right) +U^{3}=0.  \label{U3D}
\end{equation}%
At $r\rightarrow 0$, a solution of Eq. (\ref{U3D}) with an integrable
singularity is%
\begin{equation}
U_{\mathrm{3D}}(r)\approx \frac{1}{2r\sqrt{\ln \left( r/r_{0}\right) }},
\label{3D}
\end{equation}%
where $r_{0}$ is a free constant (it is relevant to mention that the 3D
stationary NLSE with the cubic term was not considered in Ref. \cite{Veron1}%
). The asymptotic form of the solution at $r\rightarrow \infty $ takes the
usual form, corresponding to the linearization of Eq. (\ref{U3D}):%
\begin{equation}
U_{\mathrm{3D}}(r)\approx Cr^{-1}\exp \left( -\sqrt{-2\omega }r\right) ,
\label{3Dexp}
\end{equation}%
with constant $C$. The matching of asymptotic forms (\ref{3D}) and (\ref%
{3Dexp}) is not considered here, as it will not determine the additional
free constant, $r_{0}$ in Eq. (\ref{3D}).

Then, it follows from Eq. (\ref{U3D}) that the \emph{exact} scaling in the $%
N(\omega )$ dependence is the same as in the 2D case, cf. Eq. (\ref{N2D}):%
\begin{equation}
N_{\mathrm{3D}}(\omega )\equiv 4\pi \int_{0}^{\infty }U^{2}(r)r^{2}dr=%
\mathrm{const}\cdot \left( -\omega \right) ^{-1/2},  \label{N3D}
\end{equation}%
hence the aVK criterion is satisfied for solutions defined by Eq. (\ref{3D})
(in terms of $\omega $, it is written as $dN/d\omega <0$), suggesting that
they may be stable.

\subsubsection{Screening of a singular attractive potential by the 3D
singular soliton}

Similar to the interpretation of the 2D model outlined above, one can
introduce, at an intermediate stage, an isotropic 3D model with a
delta-function concentrated on a sphere of small radius $\rho $, the
respective stationary equation being [cf. Eq. (\ref{U3D})]%
\begin{equation}
-\omega U-\frac{1}{2}\left( \frac{d^{2}U}{dr^{2}}+\frac{2}{r}\frac{dU}{dr}%
\right) +U^{3}-\varepsilon _{\mathrm{3D}}\delta (r-\rho )U=0.  \label{3Deps}
\end{equation}%
The solution is then looked for in the form given by Eq. (\ref{3D}) at $%
r>\rho $, and as%
\begin{equation}
U=\mathrm{const}\approx \frac{1}{2\rho \sqrt{\ln \left( r_{0}/\rho \right) }}%
~~\mathrm{at}~~r<\rho .
\end{equation}%
Then, the integration of Eq. (\ref{3Deps}) in an infinitely small vicinity
of $r=\rho $ yields the respective jump condition for the radial derivative,
cf. Eq. (\ref{rho1}):%
\begin{equation}
\left( \frac{1}{U}\frac{dU}{dr}\right) |_{r=\rho }=-2\varepsilon _{\mathrm{3D%
}}.  \label{3Drho1}
\end{equation}%
On the other hand, solution (\ref{2Drep}) yields its own value for the same
quantity:
\begin{equation}
\left( \frac{1}{U}\frac{dU}{dr}\right) |_{r=\rho }=-\rho ^{-1}.
\label{3Drho2}
\end{equation}%
Equating expressions (\ref{3Drho1}) and (\ref{3Drho2}) yields%
\begin{equation}
\varepsilon _{\mathrm{3D}}=1/\left( 2\rho \right) ,  \label{3DD}
\end{equation}%
cf. Eq. (\ref{2D}). Thus, in the limit of $\rho \rightarrow 0$, an effective
total \textquotedblleft charge" $Q_{\mathrm{3D}}$ emulating the 3D soliton
is produced by the integration of $\varepsilon _{\mathrm{3D}}$ over the
sphere of radius $\rho $:
\begin{equation}
Q_{\mathrm{3D}}=4\pi \rho ^{2}\cdot \varepsilon _{\mathrm{3D}}=2\pi \rho
\rightarrow 0.
\end{equation}%
Thus, the 3D singular solution may be interpreted as a result of screening
of the vanishingly small \textquotedblleft charge".

\subsection{A possibility of the existence of dissipative singular solitons}

As an extension of the present analysis, it is possible to look for
singular-soliton solutions in models of dissipative and $\mathcal{PT}$%
-symmetric nonlinear media. In particular, the 1D equation with a complex
coefficient in front of the septimal term,
\begin{equation}
iu_{z}=-\frac{1}{2}u_{xx}+\left( 1-i\alpha \right) |u|^{6}u,  \label{diss1D}
\end{equation}%
where $\alpha >0$ represents the strength of four-photon absorption, in
terms of the optics model (in the experiment, the absorption may be
introduced by resonant dopants). It is easy to find an exact solution to Eq.
(\ref{diss1D}) with $k=0$ and diverging integral norm (total power) [cf.
solution (\ref{-1/3}) to Eq. (\ref{NLSE})]:%
\begin{eqnarray}
u(x) &=&C_{\mathrm{1D}}\left\vert x\right\vert ^{-1/3+i\mu _{\mathrm{1D}}},~
\label{diss1Dexact} \\
C_{\mathrm{1D}}^{6} &=&\frac{5}{6\alpha }\left[ \sqrt{\left( \frac{5}{%
6\alpha }\right) ^{2}+\frac{4}{3}}-\frac{5}{6\alpha }\right] ,  \notag
\end{eqnarray}%
with the \textit{chirp coefficient},%
\begin{equation}
\mu _{\mathrm{1D}}=\sqrt{\left( \frac{5}{6\alpha }\right) ^{2}+\frac{4}{3}}-%
\frac{5}{6\alpha }.  \label{mu1D}
\end{equation}

The existence of the stationary solution in the presence of the loss term
and absence of any explicit gain may seem surprising. In fact, the nonlinear
loss is compensated by influx of power from the reservoir represented by the
diverging total norm, cf. Ref. \cite{Porras}. Aiming to construct stable
dissipative solitons with $k>0$ and a finite norm, it will be necessary to
add cubic or quintic gain to Eq. (\ref{diss1D}).

Similar to what was done above for the conservative model, the dissipative
singular soliton may be interpreted as a result of the screening of a
\textquotedblleft bare" delta-functional attractive potential, this time
with an imaginary part which represents localized gain. Accordingly, Eqs. (%
\ref{eps}) and (\ref{approx}) are replaced by
\begin{equation}
\frac{1}{2}\frac{d^{2}U}{dx^{2}}=kU+\left( 1-i\alpha \right)
U^{7}-\varepsilon \left( 1-3i\mu _{\mathrm{1D}}\right) \delta (x)U,
\end{equation}%
\begin{equation}
U_{\varepsilon }(x)\approx C_{\mathrm{1D}}\left( |x|~+\xi \right)
^{-1/3+i\mu _{\mathrm{1D}}},
\end{equation}%
where $\mu _{\mathrm{1D}}$ is the same as given by Eq. (\ref{mu1D}), and $%
\xi $ is again given by Eq. (\ref{xi}).

Similarly, one can introduce a 2D equation with a complex coefficient in
front of the quintic term, cf. Eqs. (\ref{u2D}) and (\ref{diss1D}),
\begin{equation}
iu_{z}=-\frac{1}{2}\nabla ^{2}u+\left( 1-i\alpha \right) |u|^{4}u,
\label{diss2D}
\end{equation}%
where $\alpha >0$ represents the two-photon absorption in a bulk optical
waveguide, in terms of optics. Equation (\ref{diss2D}) admits an exact
solution with $k=0$ and diverging integral power,%
\begin{eqnarray}
u(r) &=&C_{\mathrm{2D}}r^{-1/2+i\mu _{\mathrm{2D}}},~  \label{diss2Dexact} \\
C_{\mathrm{2D}}^{4} &=&\frac{1}{4\alpha }\left( \sqrt{\frac{1}{\alpha ^{2}}+1%
}-\frac{1}{\alpha }\right) ,  \notag
\end{eqnarray}%
where the chirp coefficient is
\begin{equation}
\mu _{\mathrm{2D}}=\frac{1}{2}\left( \sqrt{\frac{1}{\alpha ^{2}}+1}-\frac{1}{%
\alpha }\right) ,
\end{equation}%
cf. Eqs. (\ref{diss1Dexact}) and (\ref{mu1D}).

Finally, the three-dimensional NLSE (\ref{u3D}) with the complex coefficient
in front of the cubic term,
\begin{equation}
iu_{t}=-\frac{1}{2}\nabla ^{2}u+\left( 1-i\alpha \right) |u|^{2}u,
\label{diss3D}
\end{equation}%
admits a chirped solution with the following asymptotic form at $%
r\rightarrow 0$:%
\begin{equation}
u(r)\approx \frac{1}{2r}\left( \ln \left( \frac{r}{r_{0}}\right) \right)
^{-\left( 1-i\alpha \right) /2},  \label{diss3Dapprox}
\end{equation}%
cf. Eq. (\ref{3D}). Unlike the exact 1D and 2D solutions (\ref{diss1Dexact})
and (\ref{diss2Dexact}), Eq. (\ref{diss3Dapprox}) yields an approximate
solution in the 3D case.

Detailed consideration of the dissipative singular solitons should be a
subject for a separate work. In this connection, it is relevant to mention
that an exact singular-soliton solution is produced by an integrable $%
\mathcal{PT}$-symmetric version of NLSE with nonlocal cubic nonlinearity
\cite{Ablowitz}. However, the singularity is featured by the evolution of
the soliton in $z$, rather than by its dependence on $x$.

\section{Numerical results}

The analytical results presented above were verified and extended by
comparison with systematically generated numerical findings. In fact, the
analytical results provide a nearly complete description of the 1D, 2D, and
3D singular solitons, which is confirmed by the comparison with numerically
generated stationary solutions, as shown below. However, it is critically
important to check the stability (or instability) of the predicted states,
for which only necessary but not sufficient conditions are provided by the
analytical investigation, in the form of the aVK criterion (and its
applicability is only a conjecture). In spite of these reservations,
numerical results displayed below clearly demonstrate, by means of
systematic simulations of perturbed evolution of the stationary solutions,
that the aVK criterion provides not only necessary but also completely
sufficient conditions of the stability (for the fundamental 1D, 2D, and 3D
solitons) or instability (for the 2D singular solitons with embedded
vorticity, in the model with the quintic self-attraction).

The numerical scheme must be adjusted to the fact that, in the analytical
form, all the solutions under the consideration take infinite values at the
origin, $x=y=z=0$ or its 1D and 2D counterparts. To carry out numerical
simulations, the finite-difference scheme was used, defined on a numerical
grid constructed so that mesh points closest to the origin have coordinates
\begin{equation}
\left( x,y,z\right) =\left( \pm \Delta /2,\pm \Delta /2,\pm \Delta /2\right)
\label{Delta}
\end{equation}%
(similarly defined in the 1D and 2D cases), where $\Delta $ is the mesh
size. At these closest-to-the origin points, boundary conditions with large
but finite values of $U$ were fixed as per the asymptotically exact
analytical solutions (\ref{-1/3}), (\ref{2Drep}), and ( \ref{3D}),
respectively. Further details of the numerical scheme are given below.

\subsection{The 1D model}

\begin{figure}[h]
\begin{center}
\includegraphics[height=3.5cm]{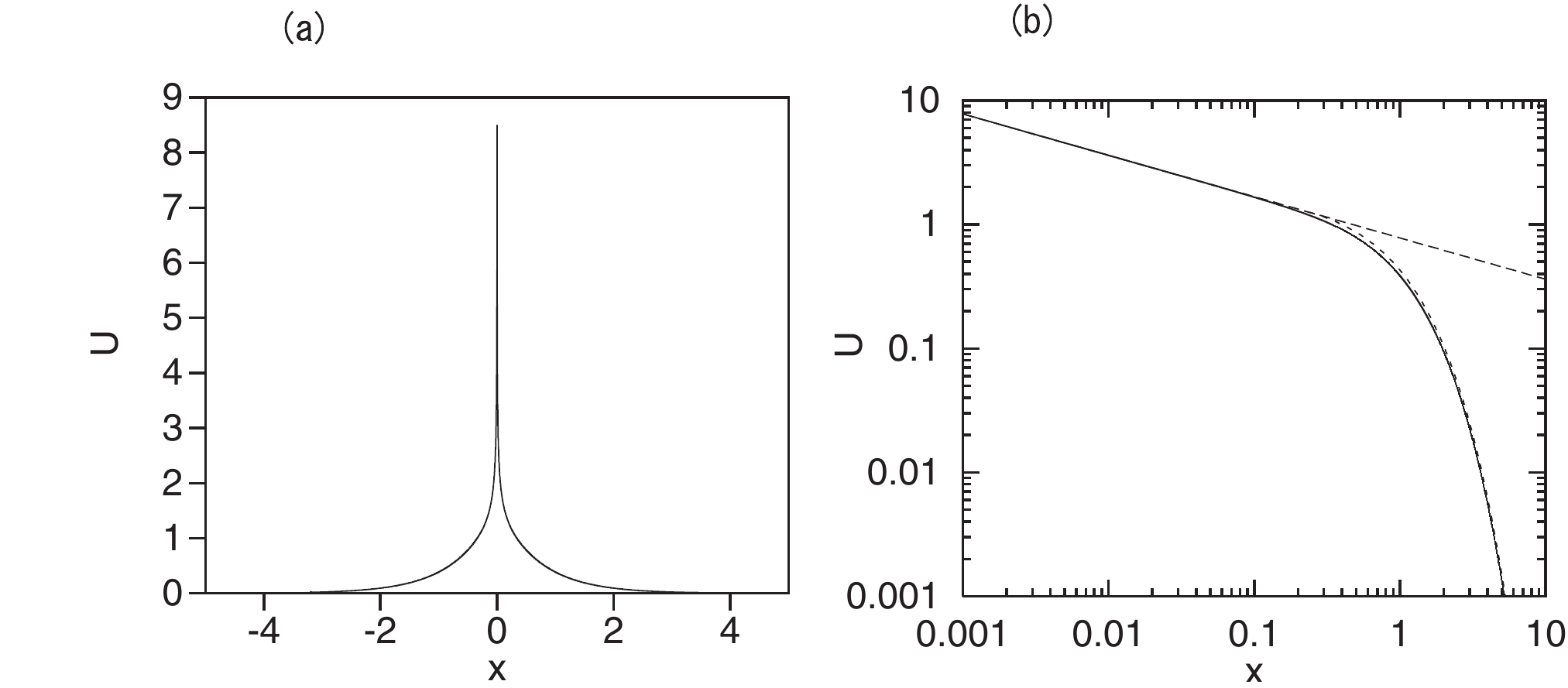}
\end{center}
\caption{(a) A singular 1D soliton with $k=1$, produced by the numerical
solution of Eq. (\protect\ref{hh}). (b) The double-logarithmic plot of the
same solution at small values of $x$ is presented by the solid curve. The
dashed line is the analytical prediction provided by Eq. (\protect\ref{-1/3}%
), and the dotted curve, which is almost indistinguishable from the solid
one, is the approximation given by Eq. (\protect\ref{match}).}
\label{f1}
\end{figure}

Singular solitons were generated as numerical solutions to Eq.~(\ref{h})
with $h=0$, written as
\begin{equation}
\frac{dU}{dx}=\sqrt{2kU^{4}+(1/2)U^{8}}.  \label{hh}
\end{equation}%
This equation was solved by means of the Runge-Kutta (RK) method with
stepsize $\Delta x=10^{-5}$. A typical profile of the numerically generated
1D singular soliton is shown in Fig. \ref{f1}(a), for $k=1$. Figure \ref{f1}%
(b) separately displays a double-logarithmic plot of the same solution for
small values of $x$, and provides its comparison to the analytical
prediction given by Eq. (\ref{-1/3}) and Eq. (\ref{match} ). The latter plot
clearly demonstrates that the analytical result is an asymptotically exact
one.

\begin{figure}[h]
\begin{center}
\includegraphics[height=3.0cm]{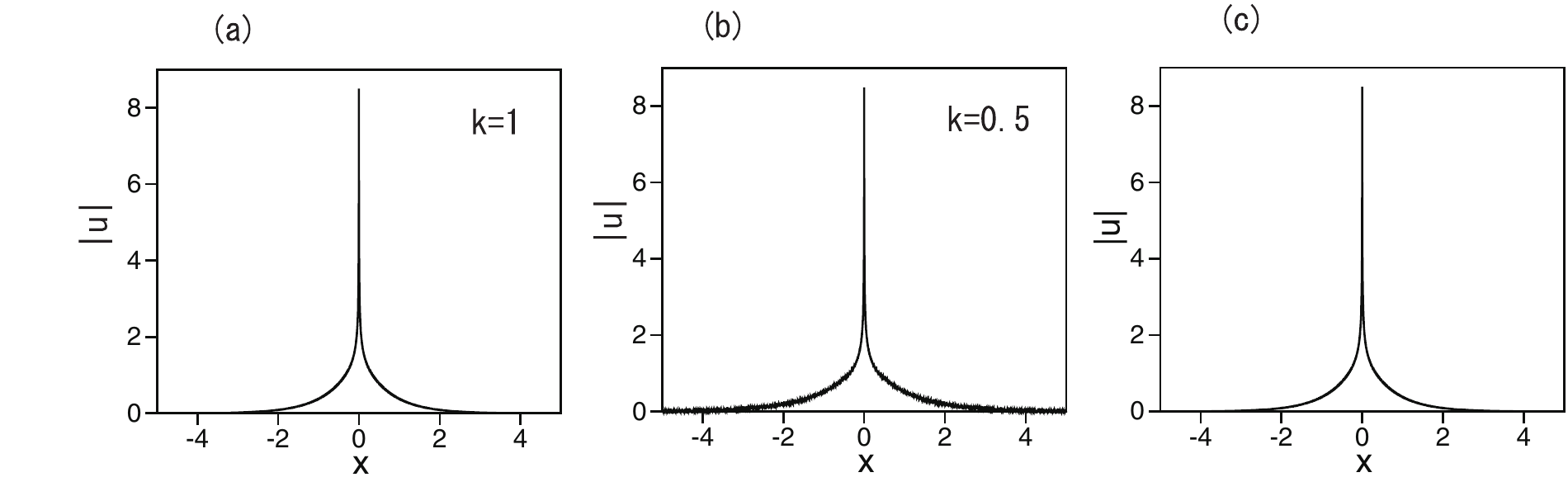}
\end{center}
\caption{Snapshots of the 1D singular solitons, $|u(x,z)|$, at $z=5$,
produced by direct simulations of Eq. (\protect\ref{NLSE}), with the input
taken as per the stationary solutions with $k=1$ (a) and $k=0.5$ (b). The
solutions keep the same shape at $z>5$. (c) The snapshot of $|u\left(
x,z=5\right) |$, produced by the simulation of Eq. (\protect\ref{NLSE}) with
boundary conditions (\protect\ref{k}) for $k=1$, initiated by the stationary
solution for the singular soliton with $k=0.5$, see the text.}
\label{f3}
\end{figure}

The stability of the entire family of the 1D singular solitons, suggested by
the fact that it satisfies the conjectured aVK criterion, is fully confirmed
by systematic simulations (the above-mentioned fixed boundary conditions at $%
x=\pm \Delta /2$ are essential for running the simulations). To illustrate
this conclusion, results of the direct simulations are displayed in Figs. %
\ref{f3}(a) and (b), for the solutions with $k=1$ and $k=0.5$. The numerical
simulations of 1D equations (\ref{U}) and (\ref{NLSE}) (with $g_{3}=g_{5}=0$%
) were performed in domain $|x|\leq 12.5$ with zero boundary conditions at $%
x=\pm 12.5$, using the split-step Fourier method based on a set of $16384$
modes,the respective mesh sizes being $\Delta x=25/16384\approx \allowbreak
1.526\times 10^{-3}$ and $\Delta z=10^{-6}$. At the central points, $x=\pm
\Delta /2$, steadily oscillating boundary values
\begin{equation}
u(\pm \Delta /2)=\exp (ikz)U(\Delta /2)  \label{k}
\end{equation}
are imposed, with $k=(3.1478/N)^{6}$ taken as per Eq. (\ref{N}), $U(\Delta
/2)$ being the value numerically evaluated by means of the RK method, as
said above. The initial conditions are provided by the stationary singular
solitons produced by the RK method. The profiles produced by the evolution
are virtually indistinguishable from the input. Simulations with small
random perturbations added to the input corroborate the stability.

Robustness of the singular solitons against large perturbations was tested
by direct simulations of Eq. (\ref{NLSE}) with boundary conditions (\ref{k}%
), with inputs strongly different from the stationary solitons corresponding
to given $k$. As an example, Fig. \ref{f3}(c) demonstrates the result for
fixed $k=1$ in Eq. (\ref{k}) with the mismatched input, taken as the
stationary singular soliton corresponding to $k=0.5$, but with the amplitude
rescaled so as to make the total norm equal to that corresponding to $k=1$.
The result, produced by the simulations at $z=5$, is indistinguishable from
the one displayed in Fig. \ref{f3}(a), which was generated by the input in
the form of the stationary soliton corresponding to the ``correct" value, $%
k=1$.

\begin{figure}[h]
\begin{center}
\includegraphics[height=7.5cm]{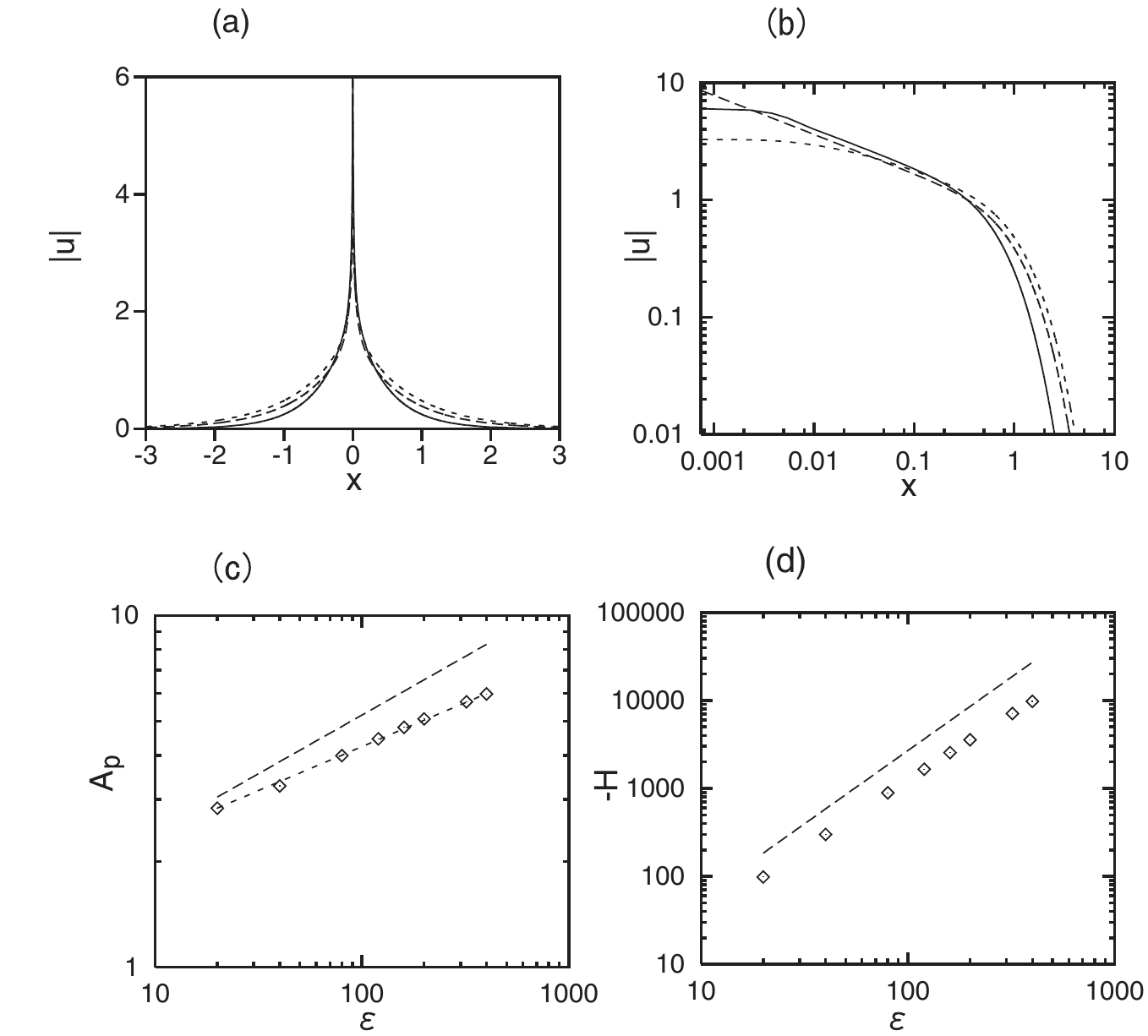}
\end{center}
\caption{ (a) The comparison of the singular-soliton solution from Fig.
\protect\ref{f1}, shown by the long-dashed profile, with numerically
generated ground-state solutions of Eq. (\protect\ref{eps}) (see details in
the text), for $\protect\varepsilon =40$ and $400$ (the short-dashed and
continuous profiles, respectively). (b) The same as in (a), but displayed in
the double-logarithmic form. (c) The chain of rhombic symbols, connected by
the short-dashed line as the guide to the eye, shows the peak amplitude of
the numerically found solutions, $A_{p}\equiv U(x=\pm \Delta /2)$, vs. $%
\protect\varepsilon $, while the long-dashed line shows the respective
analytical approximation, $A_{p}=\left( \protect\sqrt{2}\protect\varepsilon %
\right) ^{1/3}$, as given by Eqs. (\protect\ref{approx}) and (\protect\ref%
{xi}) with $x=0$. (d) The chain of symbols shows the dependence of the
Hamiltonian of numerically generated solutions on $\protect\varepsilon $,
with the dashed line showing the respective analytical approximation, given
by Eq. (\protect\ref{Heps}).}
\label{f-extra}
\end{figure}

Finally, the comparison of the singular-soliton solution with that produced
by the nonlinearity-screened attractive delta-functional potential, in the
framework of Eq. (\ref{eps}), is displayed in Fig. \ref{f-extra}. The
ground-state solution of Eq. (\ref{eps}) was obtained by means of the
imaginary-integration method, applied to the nonstationary version of Eq. (%
\ref{eps}). The approximate delta-function, $\tilde{\delta}$, was realized
on the numerical mesh by setting $\tilde{\delta}=1/\left( 4\Delta \right) $
at points $x=\pm 3\Delta /2$ and $\pm \Delta /2$, and $\tilde{\delta}=0$
elsewhere ($\Delta =25/16384\approx \allowbreak 1.526\times 10^{-3}$ and the
domain's definition, $|x|~\leq 12.5$, are the same as above). The comparison
corroborates that the singular solitons can be realized as the ground state
generated by the delta-functional potential screened by the septimal
self-repulsion. Weak divergence of the comparison at very small values of $%
|x|$ and very large $\varepsilon $ is explained by the limited accuracy of
the numerical scheme with a fixed mesh size, $\Delta $.

\subsection{The 2D model}

\begin{figure}[h]
\begin{center}
\includegraphics[height=7.5cm]{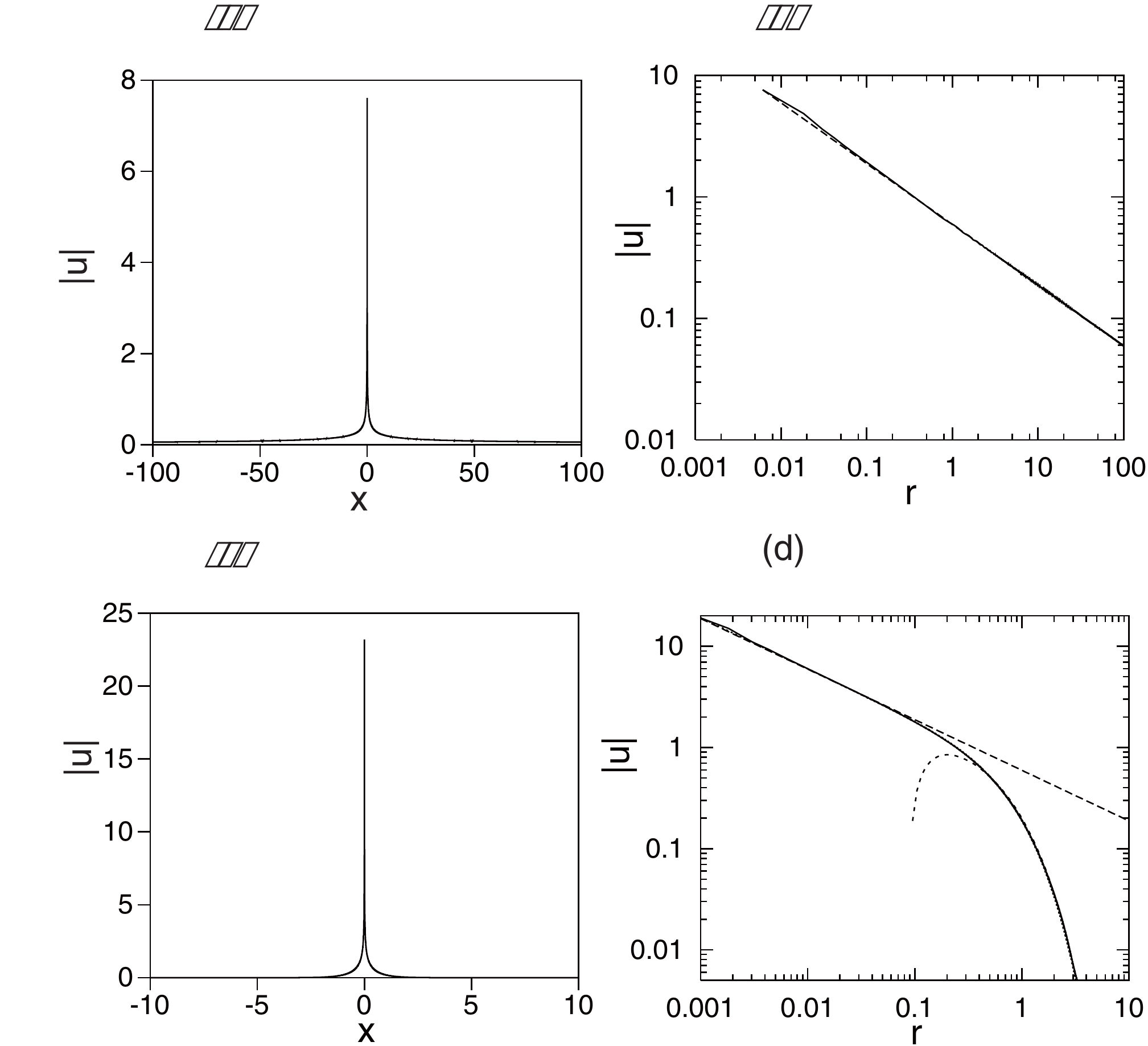}
\end{center}
\caption{(a) Snapshot of solution $|u(r,z)|$ at $z=5$, produced by direct
simulations of the 2D radial equation~(\protect\ref{u2dr}), which was
initiated by the input taken as per the exact analytical solution for $k=0$,
given by Eq. (\protect\ref{u0}) (the solution is shown by the cross section
along $y=0$). (b) The comparison of the numerical solution (the continuous
line) with the input (the dashed line) at small values of $r$, on the
double-logarithmic scale. The dashed line is the analytical expression (%
\protect\ref{u0}). (c) The snapshot of solution $|u(r,z)|$ at $z=5$ for $k=1$%
, produced by direct simulations of the 2D radial equation~(\protect\ref%
{u2dr}). (d) The snapshot of the same solution, shown on the
double-logarithmic scale. The long- and short-dashed lines represent,
respectively, the analytical expressions given by Eqs. (\protect\ref{u0})
and (\protect\ref{2Dexp}), with fitting parameter $C=0.9$.}
\label{f4a}
\end{figure}

The existence and stability of fundamental ($S=0$) 2D singular solitons,
produced by the quintic equation (\ref{u2D}) with $\sigma =+1$
(self-repulsion), has been completely corroborated by numerical solutions of
the equation. The mesh size $\Delta =25/2048\approx 1.221\times 10^{-2}$ was
used in this case. Figure \ref{f4a}(a) displays the radial profile of the
solution produced at $z=5$ by direct simulations of the radial version of
Eq. (\ref{u2D}),
\begin{equation}
i\frac{\partial u}{\partial z}=-\frac{1}{2}\left( \frac{\partial ^{2}u}{%
\partial r^{2}}+\frac{1}{r}\frac{\partial u}{\partial r}\right) +|u|^{4}u,
\label{u2dr}
\end{equation}%
with the input taken in the form of the exact analytical solution (\ref{u0})
corresponding to $k=0$. Panel \ref{f4a}(b) compares the numerical solution
with the input at small values of $r$, confirming that Eq. (\ref{u0})
provides an asymptotically exact solution for $x\rightarrow 0$. Figure \ref%
{f4a}(c) displays the radial profile of the solution of Eq. (\ref{u2D}) for $%
k=1$ at $z=5$. Panel \ref{f4a}(d) compares the numerical solution with
analytical results given by Eqs. (\ref{u0}) and~(\ref{2Dexp}) with fitting
constant $C=0.9$ (dashed and dotted lines, respectively).

\begin{figure}[h]
\begin{center}
\includegraphics[height=4.3cm]{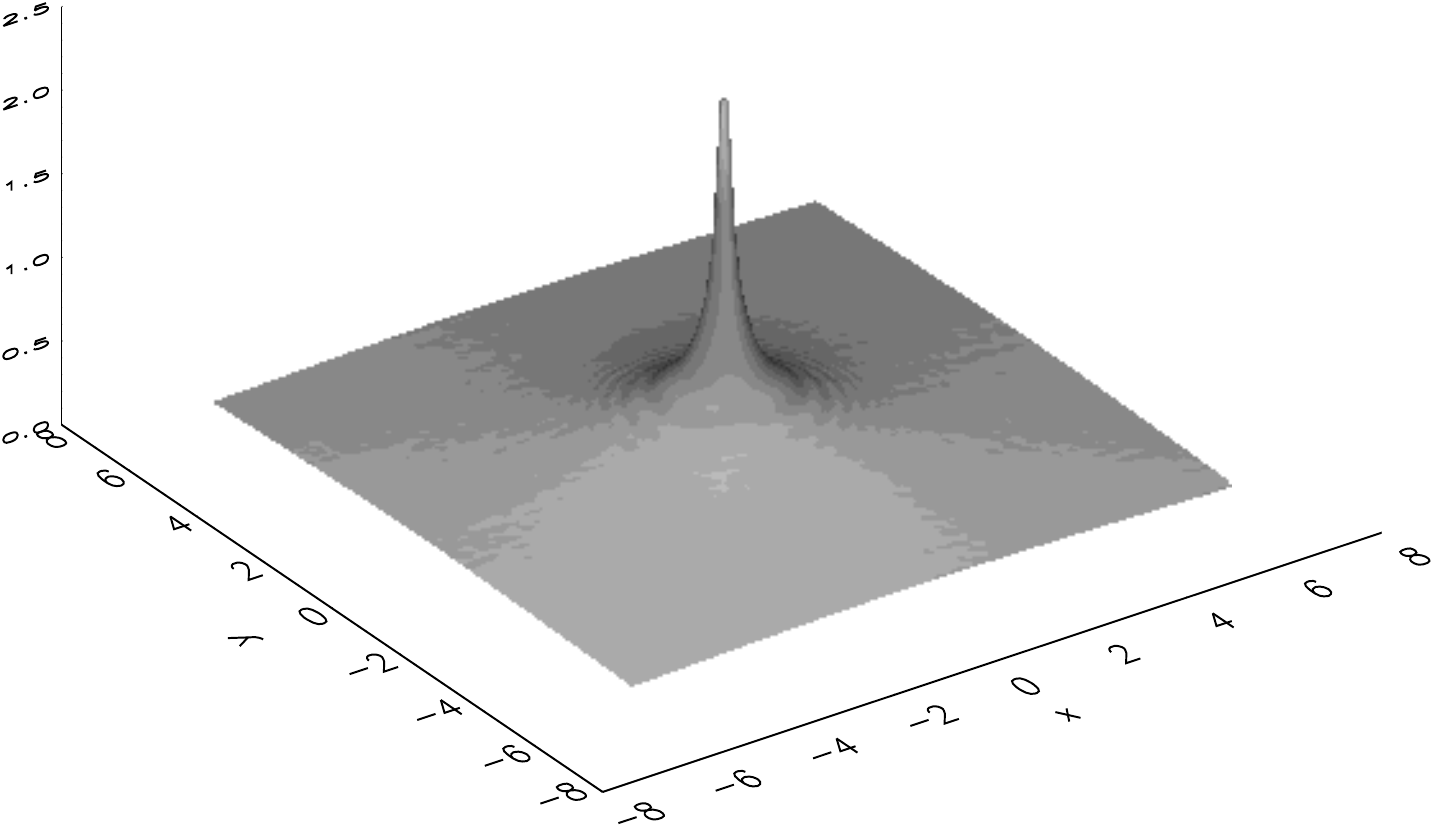}
\end{center}
\caption{The same solution as in Fig. \protect\ref{f4a}(c), but produced by
simulation of the full 2D equation~(\protect\ref{u2D}). }
\label{f4b}
\end{figure}

Further, the full stability analysis of 2D solitons must include a
possibility of azimuthal perturbations which break the axial symmetry of the
solution, which makes it necessary to run simulations of the full 2D
equation (\ref{u2D}), in addition to the simulations of the radial equation
( \ref{u2dr}). The result of the simulations, displayed in Fig. \ref{f4b},
clearly corroborates the stability of the 2D fundamental solitons against
azimuthal perturbations.

\begin{figure}[h]
\begin{center}
\includegraphics[height=3.5cm]{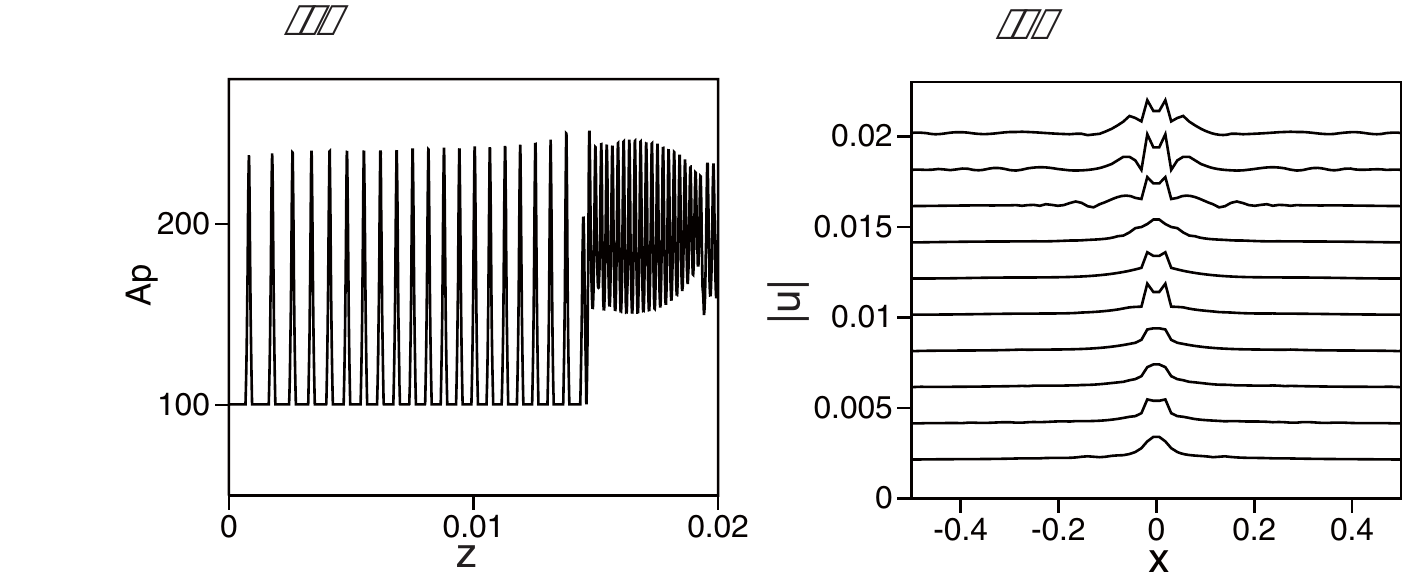}
\end{center}
\caption{(a) The evolution of the peak amplitude, $A_{p}$, of solution $%
|u(x,y,z)|$ [i.e., values of $\left\vert u\left( x,y=\pm \Delta /2\right)
\right\vert $, see Eq. (\protect\ref{Delta})], as produced by simulations of
Eq.~(\protect\ref{u2dr2}) with the quintic self-attraction, for a 2D
singular soliton with embedded vorticity $S=1$. The input is taken as per
Eq. (\protect\ref{u0S=1}). (b) The evolution of the radial profiles, $%
\left\vert u(r,z)\right\vert $, for the same solution. Numbers attached to
the vertical axis indicate values if $z$ in the course of the evolution.}
\label{f6}
\end{figure}

The above analytical consideration of the 2D singular solitons with embedded
vorticity $S\geq 1$, which are solutions of Eq. (\ref{u2D}) with $\sigma =-1$
(self-attraction), corresponding to the asymptotic form (\ref{2Dattr}), has
led to the conjecture that all the vortex solitons are unstable, as they do
not satisfy the Vakhitov-Kolokolov criterion. This prediction is confirmed,
in Fig. \ref{f6}(a), by simulations of the radial reduction of Eq. (\ref{u2D}%
) with $\sigma =-1$, for $S=1$:
\begin{equation}
i\frac{\partial u}{\partial z}=-\frac{1}{2}\left( \frac{\partial ^{2}u}{%
\partial r^{2}}+\frac{1}{r}\frac{\partial u}{\partial r}-\frac{S^{2}}{r^{2}}%
u\right) -|u|^{4}u.  \label{u2dr2}
\end{equation}%
The simulations of Eq. (\ref{u2dr2}) were performed with input
\begin{equation}
u_{0}(r)=(3/8)^{1/4}r^{-1/2},  \label{u0S=1}
\end{equation}%
taken as per Eq. (\ref{2Dattr}) with $S=1$. The instability, which may
signalize a transition to the supercritical collapse, driven by the quintic
self-attraction in 2D, is clearly seen in Fig. \ref{f6}(a). Similar
instability was found for other values of $S\neq 0$. In fact, splitting
instability of the vortex states is expected to develop still faster in the
framework of the full 2D equation (\ref{u2D}).

\subsection{The 3D model}

\begin{figure}[h]
\begin{center}
\includegraphics[height=3.5cm]{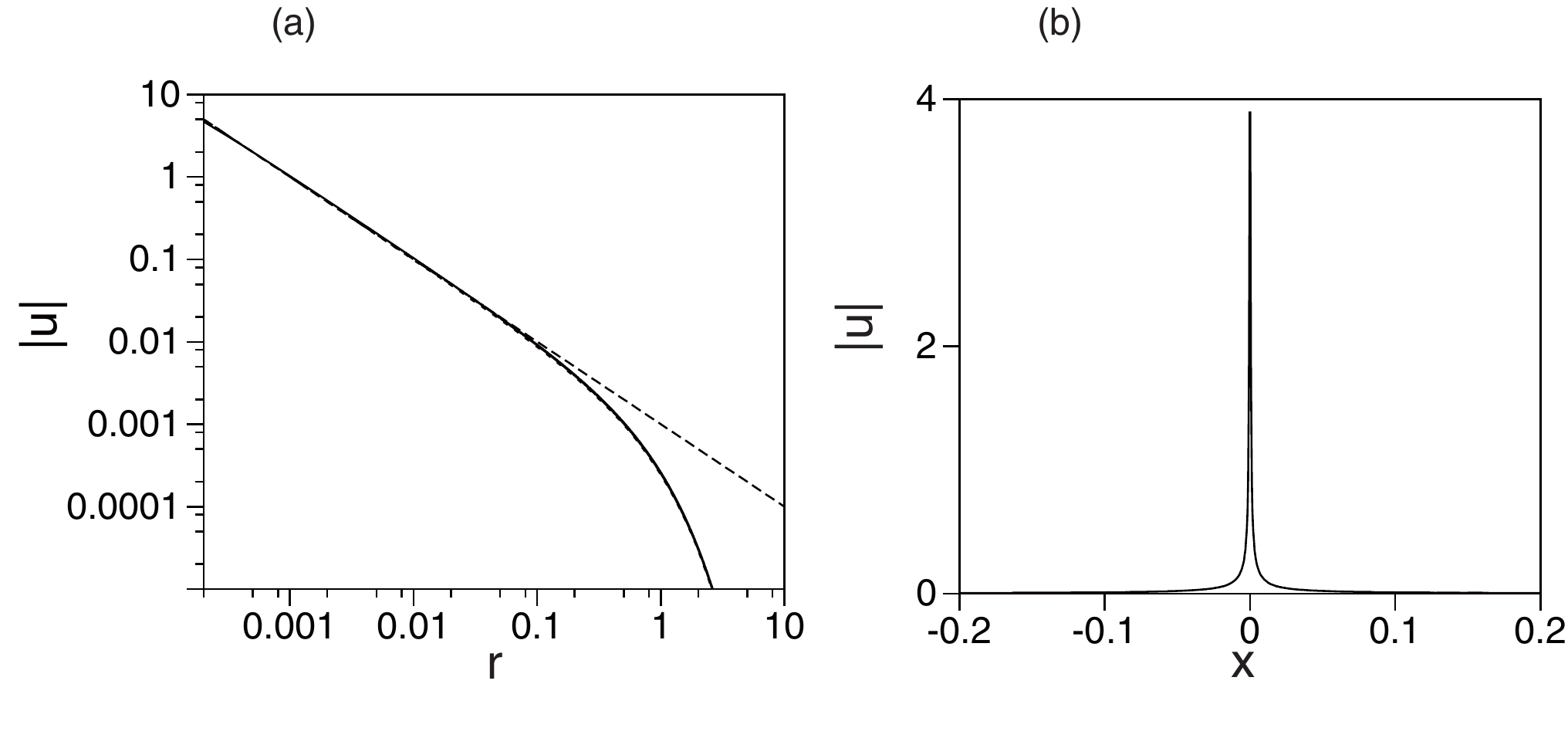}
\end{center}
\caption{(a) The comparison, on the double-logarithmic scale, of the
numerically obtained stationary solution of Eq.~(\protect\ref{U3D}) with $%
\protect\omega =-1$ (the continuous line) with the semi-analytical profiles,
$0.001\cdot r^{-1}$ (the long-dashed line) and $0.001\exp (-\protect\sqrt{2}%
r)/r$ (the short-dashed line), see the text. The difference between the
numerical solution and the latter analytical approximation is hardly
visible. (b) The radial cross section of the 3D isotropic singular soliton,
produced at $z=2$ by direct simulations of Eq.~(\protect\ref{U3Dr}), with
the initial taken configuration taken from panel (a).}
\label{f7}
\end{figure}

Numerical solutions of the 3D model were performed with the mesh size $%
\Delta =5/4096\approx \allowbreak 1.221\times 10^{-3}$. Radial profiles of
3D isotropic singular solitons were found as solutions of Eq.~(\ref{U3D}).
Formal comparison of these solutions to the analytical prediction, given by
Eq. (\ref{3D}) for $r\rightarrow 0$, is not straightforward, as the latter
expression contains a slowly varying logarithmic factor, which depends on
indefinite parameter $r_{0}$. Therefore, the comparison at small values of $%
r $ was performed with analytical profile $\mathrm{const}\cdot r^{-1}$,
where the constant was selected as the best-fit parameter. An example is
displayed in Fig. \ref{f7} (a), where analytical fits for the comparison
with the numerically found solution pertaining to $k=1$ are chosen, at small
and large $r$, as $0.001\cdot r^{-1}$ and $0.001\exp (-\sqrt{2}r)\cdot
r^{-1} $, respectively, the latter one corresponding to Eq.~(\ref{3Dexp}).
It is seen that the fit provides asymptotically exact agreement.

Finally, the stability of the 3D singular solitons was verified by direct
simulations of the radial version of Eq. (\ref{u3D}),
\begin{equation}
i\frac{\partial u}{\partial t}=-\frac{1}{2}\left( \frac{\partial ^{2}u}{%
\partial r^{2}}+\frac{2}{r}\frac{\partial u}{\partial r}\right) +|u|^{2}u=0,
\label{U3Dr}
\end{equation}%
with the initial condition taken as the stationary solution, such as the one
shown in Fig.~\ref{f7}(a). An example displayed in Fig. \ref{f7}(b) confirms
that the 3D singular solitons are stable states.

\section{Conclusion}

The objective of the present work is to demonstrate that the traditional
concept of solitons as regular self-trapped solutions, supported by the
competition of self-focusing nonlinearity and linear dispersion or
diffraction, may be extended to singular solutions with a \emph{convergent
norm}. These singular states are supported by a purely defocusing
nonlinearity, provided that it is strong enough, \textit{viz}., septimal,
quintic, and usual cubic terms, in the 1D, 2D, and 3D settings, respectively
(the possibility of such singularities, produced by the stationary 1D and 2D
equations, was first reported in Ref. \cite{Veron1}). Such counter-intuitive
solutions may be naturally interpreted as describing a result of complete
screening of a delta-functional attractive potential by the repulsive
nonlinearity, the respective \textquotedblleft charge" (full strength of the
attractive potential) being infinite in 1D, finite in 2D, and vanishing in
3D. The required septimal and quintic self-repulsive nonlinearity can be
realized in planar and bulk colloidal waveguides for light beams, while the
3D equation with the cubic self-repulsion is the usual GPE\
(Gross-Pitaevskii equation) for BEC. The realization of the singular
solitons in terms of the screened delta-functional attractive potential
suggests a possibility of their experimental creation. In particular, a
narrow stripe carrying a high value of the refractive index, embedded in a
planar solid-colloidal planar waveguide, may be used for the making of the
effectively 1D soliton. The comparison with numerical results confirms that
the analytical results provide accurate approximations for the singular
solitons at small and large distances from the origin. Stability of the
singular states is exactly predicted by the conjectured aVK
(anti-Vakhitov-Kolokolov) criterion. In addition to the fundamental
solitons, the 2D model with the quintic self-attraction admits singular
solitons with embedded vorticity, but they are completely unstable. The
possibility of the existence of relevant solutions in the form of
dissipative singular solitons, in the 1D, 2D, and 3D models with a complex
coefficient in front of the nonlinear term, is briefly considered too.

As an extension of the present analysis, it may be relevant to consider
interactions between stable singular solitons.

\section*{Acknowledgments}

The work of H.S. is supported by a Grant-in-Aid for Scientific Research (No.
18K03462) from the Ministry of Education, Culture, Sports, Science and
Technology of Japan. B.A.M. appreciates support provided the Israel Science
Foundation, through grant No. 1287/17, and hospitality of the
Interdisciplinary Graduate School of Engineering Sciences at the Kyushu
University.

\end{document}